\documentclass[runningheads]{llncs}

\usepackage{graphicx}
\usepackage{epstopdf}
\usepackage{color}
\usepackage{url}
\usepackage{booktabs}
\usepackage{textgreek}

\definecolor{myred}{rgb}{.8,.0,.0}

\begin{document}
\title{Multi-task Ensembles with Crowdsourced Features Improve Skin Lesion Diagnosis}
\titlerunning{Crowdsourced Features Improve Skin Lesion Diagnosis}

\author{Ralf Raumanns\inst{1,2} \and
Elif K Contar\inst{3} \and
Gerard Schouten\inst{1,2} \and \\
Veronika Cheplygina\inst{2}}

\authorrunning{R. Raumanns et al.}
\institute{Fontys University of Applied Science, Eindhoven, The Netherlands \and
Eindhoven University of Technology, Eindhoven, The Netherlands \and
Middle East Technical University, Ankara, Turkey}

\maketitle

\begin{abstract}
Machine learning has a recognised need for large amounts of annotated data. Due to the high cost of expert annotations, crowdsourcing, where non-experts are asked to label or outline images, has been proposed as an alternative. Although many promising results are reported, the quality of diagnostic crowdsourced labels is still unclear. We propose to address this by instead asking the crowd about visual features of the images, which can be provided more intuitively, and by using these features in a multi-task learning framework through ensemble strategies. We compare our proposed approach to a baseline model with a set of 2000 skin lesions from the ISIC 2017 challenge dataset. The baseline model only predicts a binary label from the skin lesion image, while our multi-task model also predicts one of the following features: asymmetry of the lesion, border irregularity and color. We show that multi-task models with individual crowdsourced features have limited effect on the model, but when combined in an ensembles, leads to improved generalisation. The area under the receiver operating characteristic curve is 0.794 for the baseline model and 0.811 and 0.808 for multi-task ensembles respectively. Finally, we discuss the findings, identify some limitations and recommend directions for further research. The code of the models is available at \url{https://github.com/raumannsr/hints_crowd}. 

\keywords{Crowdsourcing \and Multi-task learning \and Skin cancer \and Ensembles}
\end{abstract}

\section{Introduction}
Machine learning (ML) is a powerful tool for diagnosing medical images, even achieving success rates surpassing experts~\cite{Ehteshami_Bejnordi2017-gs,Rajpurkar2017-hk} However, ML is not yet widely used in healthcare, partly due to the lack of annotated data, which is time-consuming and costly to acquire. Recently, crowdsourcing -- annotating images by a crowd of non-experts -- has been proposed as an alternative, with some promising results, for example~\cite{ONeil2017,Cheplygina2016,maier2014can}. In particular, studies which rely on segmenting structures have been successful~\cite{orting2019survey}.

There are two problems with current crowdsourcing studies. One is that crowds are not able to provide diagnostic labels for images when a lot of expertise is required. For example, in \cite{albarqouni2016aggnet} crowds alone do not provide labels of sufficient quality when asked to detect mitosis in histopathology images. Another problem is that crowd annotations need to be collected for previously unseen images, which is not a desirable situation in practice. 

We propose to address both problems by multi-task learning with crowdsourced visual features. Instead of asking the crowds to provide a diagnostic label for an image, we query visual features, like color and shape. These can be provided more intuitively, but could potentially still be informative to machine learning algorithms. To address the problem of requiring crowd annotation at test time, we propose a multi-task learning setup where the crowdsourced features are only needed at the output side of the network, and thus only required when training the algorithm. In previous work multi-task learning has shown good results when combined with features provided by experts \cite{Murthy2017-lq,Hussein2017-xs} or automatically extracted by image processing methods \cite{Dhungel2017-in}. In this study we investigate for the first time  whether multi-task learning is also beneficial for crowdsourced visual features. Additionally, we investigate the effects of combining these crowdsourced features in an ensemble. 

We specifically investigate the problem of skin lesion diagnosis from dermoscopy images. We collected crowd features for asymmetry, border irregularity and color of the lesions and added these features to a baseline model with binary classification. We show that these features have limited effect when used individually, but improve performance of the baseline when used in an ensemble. This suggests that crowds can provide relevant, but complementary information to ground truth diagnostic labels.

\section{Methods} 

\subsection{Data}
\begin{figure}
    \centering
    \includegraphics[width=0.7\textwidth]{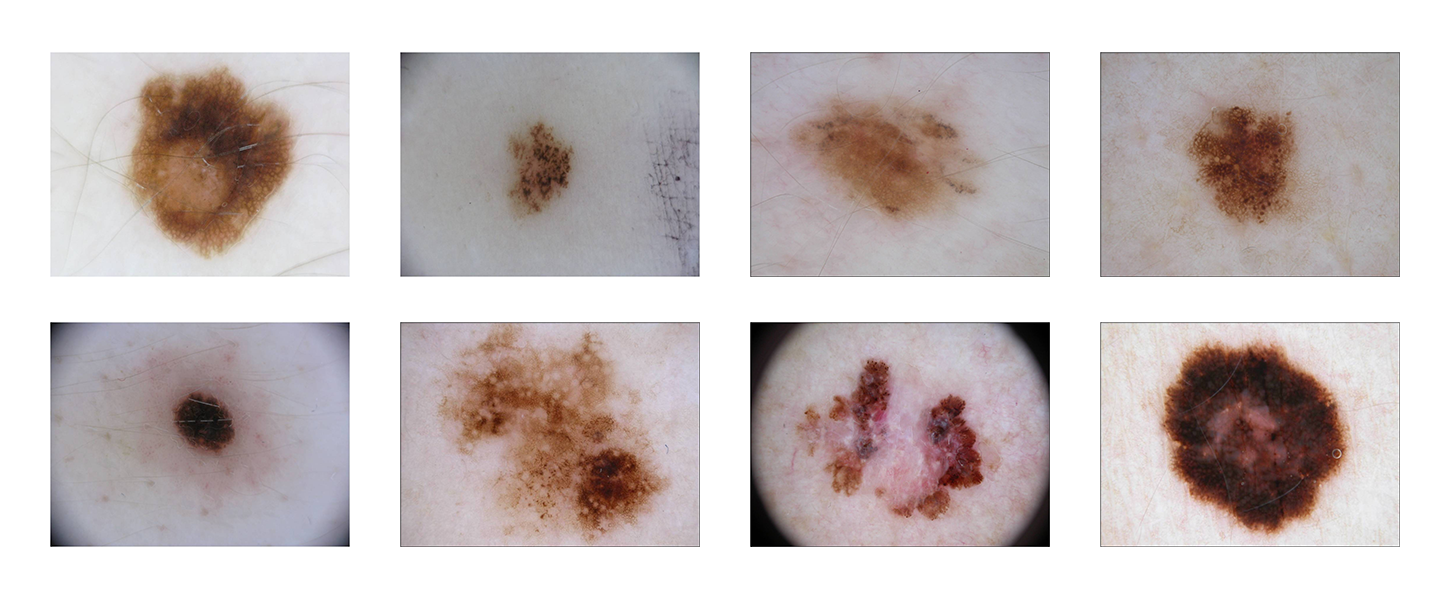}
    \caption{Examples of benign (top row) and malignant (bottom row) skin lesions from the ISIC 2017 challenge}
    \label{fig1}
\end{figure}

The dataset was extracted from the open challenge dataset of ISIC 2017 challenge \cite{Codella2017-rd}, where one of the goals is to classify the lesion as melanoma or not. Examples of the skin lesions are shown in Fig.~\ref{fig1}. 

The 2000 skin lesion images from the ISIC 2017 training data were used for training, validation and test purposes in our study. The dataset contains three classes of skin lesions: melanoma (374 lesions), seborrheic keratosis (254 lesions) and nevus (1372 lesions). The images with melanoma and seborrheic keratosis classes were combined and labeled malignant, the nevus images were labeled benign. For each malignant skin lesion there are about two benign lesions in the dataset.

\begin{figure}
    \centering
    \includegraphics[width=\textwidth]{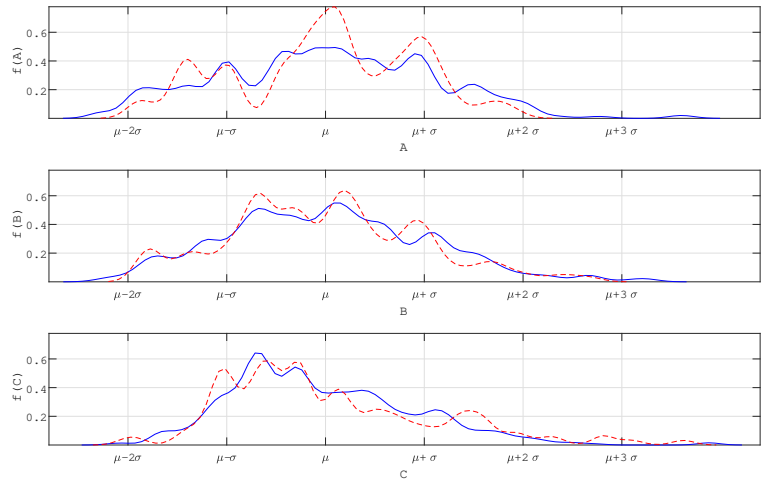}
    \caption{Probability densities f(A), f(B) and f(C) derived from the normalized asymmetry A, border B and color C measurements. The blue solid line shows density for benign lesions, the red broken line for malignant ones. The mean {\textmu} and standard deviation {\textsigma} are -0.002 and 0.766 for A, -0.010 and 0.759 for B and 0.024 and 0.897 for C measurements.}
    \label{pdf}
\end{figure}

Crowd features were collected from undergraduate students following a similar protocol as in \cite{Cheplygina2018-ee}. The students assessed the ABC features \cite{Abbasi2004-og} used by dermatologists: A for asymmetrical shape, B for border irregularity and C for color of the assessed lesion. In total 745 lesions were annotated, where each image was annotated by 3 to 6 students. Each group assessed a different set of skin lesions. From the full set of 745 skin lesions, 744 have asymmetry features, 745 have border features and 445 have color features. Before proceeding with the experiments, we normalized the measurements per student (resulting in standardized data with mean 0 and standard deviation 1) and then averaged these standardized measurements per lesion.The probability densities for the ABC features, derived from the normalized measurements, are shown in Fig.~\ref{pdf}. The differences between the probability densities for malignant and benign lesions suggest that the annotation score is informative for a lesion's label.
{}

As preprocessing we applied augmentation, i.e. random rotation with a maximum of 180 degrees, horizontal and vertical flipping, a width and height image adjustment of 10\%, a shear range of 0.2 degrees in counterclockwise direction and a random channel shift range of maximum 20 affecting the image brightness. After that the resulting images are rescaled to (384,384) pixels. 

\subsection{Models}

\begin{figure}
    \centering
    \includegraphics[width=\textwidth]{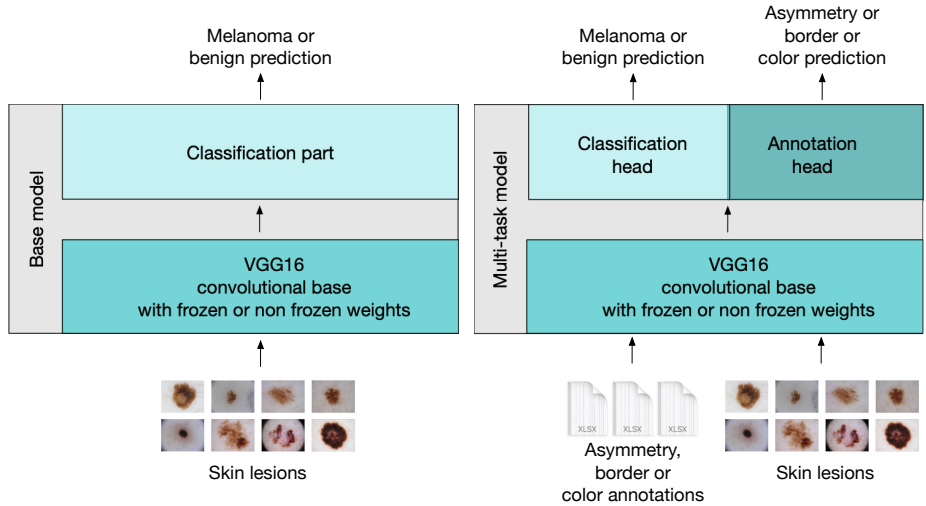}
    \caption{Architecture of baseline and multi-task models. All models built on top of the VGG16 convolutional base.}
    \label{fig2}
\end{figure}

We used two models in our experiments: a baseline model and a multi-task model (see Fig.~\ref{fig2}). Both models were built on a convolutional base and extended by adding specific layers. As encoder we used the VGG16 \cite{Simonyan2014-rs}  convolutional base. For this base, containing a series of pooling and convolutional layers, we applied fixed pre-trained ImageNet weights. We have trained the baseline and multi-task models in two ways: a) freeze the convolutional base and train the rest of the layers and b) train all layers including the convolutional base. 

Furthermore, we used two different ensembles of the multi-task models with asymmetry, border and color features. The first ensemble used equal weights for each of the three multi-task models, while the second ensemble used a weighted averaging strategy.

{}

The baseline model extended the convolutional base with two fully-connected layers with a sigmoid activation function. During training class weights were used to pay more attention to samples from the under-represented class. Cross-entropy is used as loss function. All choices for the baseline model were made based on only the training and validation set. Fine-tuning the baseline model was done until the performance on the validation dataset was within the performance of the ISIC challenge.  

The multi-task model extended the convolutional base with three fully connected layers. The model has two outputs with different network heads: one head is the classification output, the other represents the visual characteristic (Asymmetry, Border or Color). A customized mean squared error loss together with a last layer linear mapping is used for the annotation-regression task.  We used a vector in which we store per lesion whether or not a crowdsourced feature is available. This vector is used in a custom loss function that calculates the mean squared error. In case no feature is available, it has no effect on the weights and the error is not propagated.
For the binary classification task, again a cross-entropy loss and sigmoid activation function of the nodes are used. The contribution of the different losses are equal. The resulting loss values are summed and minimised during network training.

The ensembles are based on the predictions of the three available multi-task models (asymmetry, border and color) and two ensemble strategies: averaging and optimized weighted averaging.

\subsection{Experimental setup}

We compared six models with two variants for each model: a variant with a frozen convolutional layer and fully trained additional layers, and a variant in which all layers were trained. The six models were:
\begin{itemize}
    \item Baseline model with binary classification
    \item Multi-task model with the Asymmetry feature
    \item Multi-task model with the Border feature
    \item Multi-task model with the Color feature
    \item Ensemble with averaging the multi-task models predictions 
    \item Ensemble with optimized weighted averaging the multi-task models predictions
\end{itemize}

We used 5-fold cross-validation, with the dataset split in a stratified fashion, keeping the malignant benign ratio equal over the training, validation and test subset. More specifically 70\% of the dataset is used as the train subset (1400 lesions), 17.5\% as the validation subset (350 lesions), leaving 12.5\% as the test subset (250 lesions). The percentage of malignant lesions for each of the three subsets is approximately 32\%.

We trained both models iterating over 30 epochs with a batch size of 20 using the default back propagation algorithm RMSprop\cite{Tieleman2012-kp} as the optimizer, with a learning rate of $2.0\mathrm{e}{-5}$.

We performed an ensemble of the predictions of multi-task models through two strategies. In the first strategy, called averaging, the lesion's classification was based on the predictions of the three multi-task models, with the prediction of each model having an equal weight (a third). In the second strategy the classification was based on optimized weights for the prediction of each multi-task model. We used the differential evolution global optimization algorithm \cite{Storn1997-mo}, with a tolerance of $1.0\mathrm{e}{-7}$ and maximum of 1000 iterations, to find the optimized weights.

We compared the average of the area under the receiver operating characteristic curve (AUC) of the baseline model to the average AUC scores of the different multi-task models and ensembles. The average AUC score was calculated per experiment taking the average of the AUC score of each fold.

We implemented our deep learning models in Keras using the TensorFlow backend \cite{Geron2019-rs}. The code is available on Github: \url{https://github.com/raumannsr/hints_crowd}. 

\section{Results}

The AUC results obtained with the six models and two variants (frozen and not frozen convolutional layer) are shown in Table~\ref{tab:auc_results}.

\begin{table}[]
    \centering
    \begin{tabular}{c|c|c}
    \hline
    Model & AUC (mean $\pm$ std) Frozen & AUC (mean $\pm$ std) Non Frozen \\ [0.5ex]
    \hline
    
    Baseline & $0.753 \pm 0.025$ & $0.794 \pm 0.030$ \\
    
    Asymmetry & $0.750 \pm 0.025$ & $0.799 \pm 0.027$ \\
    Border &  $0.755 \pm 0.029$ & $0.782 \pm 0.020$ \\
    Color & $0.756 \pm 0.025$ & $0.770 \pm 0.016$ \\
  
    Averaging & $0.752 \pm 0.027$ & $0.808 \pm 0.025$ \\
    Optimized weighted averaging & $0.753 \pm 0.025$ & $0.811 \pm 0.028$ \\
 
    \hline
    \end{tabular}
    \caption{AUC of the six different models in two variants: frozen and non frozen convolutional base.}
    \label{tab:auc_results}
\end{table}

\begin{figure}
    \centering
    \includegraphics[width=0.9\textwidth]{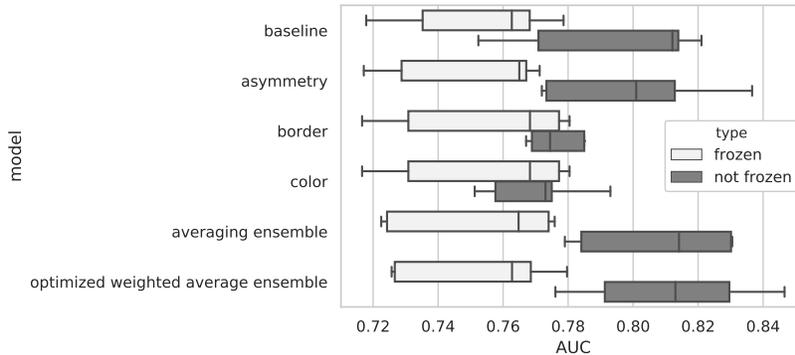}
    \caption{AUC of the six different models: baseline model, multi-task with asymmetry, multi-task with border, multi-task with color, ensemble with averaging and ensemble with optimized weighted averaging}
    \label{fig3}
\end{figure}

As can be seen in Table~\ref{tab:auc_results} and Fig.~\ref{fig3} the AUC values for the frozen variant are close together. We observed that the AUC values of the non frozen variant outperformed the frozen variant. The results of the non frozen ensemble strategies show an improved generalisation. The ensemble with optimized weighted averaging outperformed the averaging ensemble by achieving an AUC of $0.811 \pm 0.028$. Note that these numbers are somewhat lower than the ISIC 2017 challenge performances, however, we did not focus on optimizing performance but rather illustrating the added value of the multi-task ensembles approach.

{}

\section{Discussion and conclusions}

We addressed two problems in this paper. One is that crowds are not able to provide diagnostic labels when domain specific expertise is required. Another problem is that crowd features need to be collected for previously unseen images. The aim of this paper was to validate the proposed solution to tackle both problems with a multi-task learning setup using crowdsourced features. We showed that crowdsourced visual features (solely used when training the algorithm) in a multi-task learning set-up can improve the baseline model performance. This result supports the idea that crowdsourced features used at the output side of a multi-task network are a valuable alternative to possibly inaccurate diagnostic crowdsourced labels. Our work provided new insights into the potential of crowd features. Individual crowdsourced features have limited effect on the model, but when combined in an ensemble, have a bigger effect on model performance.

We sought to investigate the effect of adding crowdsourced features in a multi-task learning setting and did not focus on the absolute performance of the model. We first developed the baseline model until the performance on the test set was within the range of performances of the ISIC challenge. We did not try to optimise the model further to achieve the highest performance. For example, we have cropped the images to the same size for convenience, but this could lead to decreased the performance. To further validate the usefulness of crowdsourced features and their ensembles, it would be straightforward to extend other types of models with an additional output. 

We hypothesize that pre-training the network with domain specific data (instead of ImageNet) would further increase the AUC. As a next step we will validate two more strategies. The first strategy is to train the network on a segmentation task (using the available segmentations) and then fine-tune the network on the diagnosis and the crowdsourced features. The second strategy is to first train the network on the crowdsourced features and then fine-tune on the diagnosis. 

We averaged the normalized crowdsourced features and gave the crowd output the same weight as the true label of the lesion. This might not be the best choice and other combining strategies could be investigated. In particular, adding each annotator as an individual output could be of interest. Cheplygina et al \cite{Cheplygina2018-ee} showed that the disagreement of annotators about the score - as measured by the standard deviation of scores for a particular lesion - was informative for that lesion's label. Exploiting the similarities an differences between the annotators in our study could therefore further increase performance. 

There are some limitations with the visual assessment of the crowd. First of all, the annotators were not blinded to the true labels of the skin lesions due to the public nature of the dataset. However, if the visual assessments were perfectly correlated with the true labels, multi-task learning would not have been helpful. The information that the crowd provides therefore appears to be complementary to what is already present in the data. The next step would be to validate the method with other types of crowdsourced features like Amazon Mechanical Turk or a similar platform.

\subsection{Negative results}

In an earlier version of the paper (available on arXiV, \url{https://arxiv.org/abs/2004.14745}) we reported that multi-task models with a single feature had a substantial improvement over the baseline model. Upon closer inspection, it turned out that this result was observed due to a bug in Keras. Adding the VGG16 model as first layer and weight freezing the newly added layer does not work as advertised - the weights do continue to be updated due to the bug. As such, we compared a baseline model with frozen weights, with multi-task models with unfrozen weights, which is an unfair comparison. This issue (\url{https://github.com/keras-team/keras/issues/8443}) was partly fixed by Keras, but still did not work for the multi-task models. A workaround used for this paper is setting all inner layers of the convolutional base as non-trainable. When factoring this out, there wasn't a clear gain in performance, however, ensembles still proved to be useful.  

\section*{Acknowledgments}
We would like to acknowledge all students who have contributed to this project with image annotation and/or code. We gratefully acknowledge financial support from the Netherlands Organization for Scientific Research (NWO), grant no. 023.014.010.

\bibliographystyle{splncs04}
\bibliography{refs_ralf,refs_veronika}
\end{document}